\documentclass[english]{article}
\usepackage[T1]{fontenc}
\usepackage[latin1]{inputenc}
\usepackage{geometry}
 \usepackage{graphicx}
\usepackage{color}

\makeatletter

\providecommand{\tabularnewline}{\\}

\usepackage{babel}
\makeatother
\begin{document}
\begin{center}
\textbf{\large On the Synthesis of Sequential} \textbf{\textcolor{black}{\large Reversible}}
\textbf{\large Circuit }
\par\end{center}{\large \par}

\begin{center}
Anindita Banerjee%
\footnote{aninditabanerjee2000@yahoo.com%
} and Anirban Pathak%
\footnote{anirban.pathak@jiit.ac.in%
}
\par\end{center}

\begin{center}
Department of Physics and Material Science and Engineering, JIIT University,
A-10, Sector-62, Noida, UP-201307, India.
\par\end{center}

\begin{abstract}
\textcolor{black}{Reversible} circuits for SR flip flop, JK flip flop,
D flip flop, T flip flop, Master Slave D flip flop and Master Slave
JK flip flop have been provided with three different logical approaches.
All the circuits have been optimized with the help of existing local
optimization algorithms (e.g. template matching, moving rule and deletion
rule) and the optimized sequential circuits have been compared with
the earlier proposals for the same. It has been shown that the present
proposals have lower gate complexities and lower number of garbage
bits compared to the earlier proposals. It has also been shown that
the advantage in gate count obtained in some of the earlier proposals
by introduction of New gates is an \textcolor{black}{artifact} and
if it is allowed then every circuit block (unless there is a measurement)
can be reduced to a single gate. Further, it is shown that a reversible
flip flop can be constructed even without a feedback. In this context,
some important conceptual issues related to the designing and optimization
of sequential reversible circuits have also been addressed.
\end{abstract}

\section{Introduction}

Landauer's principle {[}\ref{landuer1}] states that any logically
irreversible operation on information, such as the erasure of a bit
or the merging of two computation paths, is always associated with
an increase of entropy of the non-information bearing degrees of freedom
of the information processing apparatus or its environment {[}\ref{Bennet}]
and consequently each bit of lost information will lead to the release
of at least $kTln2$ amount of heat. But it is well known that most
of the commonly used classical gates (except NOT and IDENTITY) are
irreversible and they erase at least one bit of information in every
operation. Thus the irreversible logic gates will always release some
heat energy. On the other hand, Moore's law states that the number
of transistors in a chip gets doubled in every 18 months. Therefore,
if we continue to design chips with the help of conventional irreversible
logic gates then the lower limit of power loss will continue to increase.
This led to the idea of reversible computation and reversible logic.
Since quantum mechanics is essentially reversible, quantum mechanical
processes appeared as good candidate to construct reversible gates
and these gates are known as quantum gates. When we glue some of these
gates we obtain a quantum circuit. After introduction of the idea
of quantum computation it has already been seen that there exist some
quantum algorithms {[}\ref{the:Nielsen and Chuang}] which works much
faster than their classical counter parts, there exists infinitely
secured quantum cryptographic protocol {[}\ref{crypto1}, \ref{crypto2}];
there exists protocol for quantum teleportation {[}\ref{the:cerf-teleportation}]
and all these processes which establish quantum computing as a superior
future technology involves quantum circuits and quantum gates. It
has also been seen that it is possible to design classical reversible
gates and classical reversible circuits but since they can not handle
superposition of states (qubit) they just form special cases of quantum
circuit or a subset of the set of the quantum circuits. But from the
construction point of view they are easy to build. Keeping this background
in mind it is reasonable to state that all novel applications of reversible/quantum
computation essentially involve reversible/quantum circuits.

A considerable amount of work has already been done in the field of
designing and optimization of reversible combinatorial circuit {[}\ref{Maslov1}-\ref{AD VOS}].
But the designing aspect of reversible sequential circuit is not yet
studied rigorously. This is because of the fact that feedback in a
reversible circuit can not be visualized in the usual sense in which
feedback is visualized in a conventional irreversible circuit. This
issue was first addressed by Toffoli {[}\ref{toffoli}]. In {[}\ref{toffoli}]
he had shown that the reversible sequential circuits can be constructed
provided the transition function of the circuit block without the
feedback loop is unitary. His ideas on the sequential reversible circuit
had further strengthen in his pioneering work on conservative logic
{[}\ref{the:Toffoli-conservative logic}]. Later on some efforts have
been made to construct reversible sequential circuit {[}\ref{picton}-\ref{thapiyal2}]
. All these efforts are concentrated on the designing of various flip
flops because of the fact that the flip flops are the basic building
block of the memory element of a computer and if one wishes to build
a reversible classical computer then these designs will play a crucial
role. But several conceptual issues related to designing and optimization
of sequential circuits are not addressed till now. In the next section
we address those conceptual issues related to the feedback and the
choice of gate library. In section 3 we have described different architectures
for flip flops. In section 4 we have compared the circuit architectures
proposed in the present work with the existing circuit architectures
for the same. Finally we conclude the work in section 5.

\section{Conceptual issues related to reversible circuit }

To provide a systematic protocol for designing reversible sequential
circuit and to compare the proposed circuit architectures with the
existing architectures we need to address certain conceptual issues
related to reversible circuit designing. To be precise, conceptual
issues related to feedback, choices of gate library and approximate
optimization (local optimization) techniques will be addressed in
the following subsections.

\subsection{Feedback in a reversible circuit}
\begin{figure*}[h]
\centering \scalebox{0.4}{\includegraphics{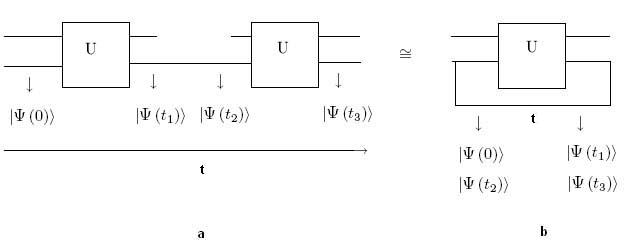}} \caption{Circuit a and b are equivalent but the idea
of time axis is not valid in b. Here $\left|\Psi(t)\right\rangle $ is the product state at time t and
$t_{3}$>$t{}_{2}$>$t_{1}$>0 }\label{fig1}.
\end{figure*}
It is widely believed that feedback is not allowed in a reversible circuit {[}\ref{vasudevan}]. This is true if
we consider feedback in a similar fashion as it is dealt in classical irreversible logic. The objection against
feedback is twofold. Firstly, merging of two computational paths is not allowed in a reversible circuit and
secondly, time axis goes from left to right in a reversible circuit (as shown in Fig. 1a). Thus if we need to
follow the same notion of time axis in a reversible sequential circuit then feedback will essentially mean a
journey in negative time axis or existence of time machine. This is against the notion of physical reality. But
these strong objections against feedback in reversible circuit can be circumvented by establishing the
equivalence between the circuits in Fig. 1a and Fig. 1b. To be precise, the feedback loop shown in Fig. 1b is
only in space not in time. Therefore, the circuit in Fig. 1b is equivalent to a cascaded circuit in time axis
(see Fig. 1a). Thus the usual notion of time axis is not valid in reversible sequential circuit (i.e. in a
circuit having spatial feedback loop similar to one shown in Fig. 1b). Further, since the circuit in Fig. 1b is
equivalent to the cascade shown in Fig 1a, there is no merging of computational paths and consequently there
would not be any loss of energy provided U is unitary. This conclusion coincides with the Toffoli's idea
{[}\ref{toffoli}] of unitary transition function. Now if we follow, this notion of feedback, then to establish
the reversibility of the architecture it would be sufficient to establish the unitarity of U. Here we would also
like to note that in this restricted notion of spatial feedback we can not allow any arbitrary feedback loop. An
allowed loop has to be reducible to structure shown in Fig. 1b.

\subsection{How to design the circuit}
\begin{figure*}[h]
\centering
 \scalebox{0.4}{\includegraphics{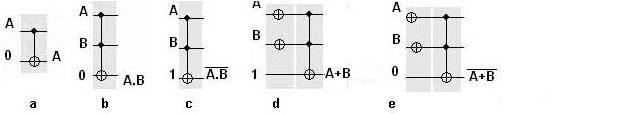}}
\caption{Realization of classical irreversible operations with the help of reversible gates: a) COPY gate, b)
AND gate, c) NAND gate, d) OR gate, e) NOR gate }\label{Fig2}
\end{figure*}

In the previous subsection we have shown that in order to design a
reversible sequential circuit we have to design U (in Fig. 1b) as
unitary. \textcolor{red}{}Now if we know the truth table of ~U and
wish to decompose U in terms of finite number of logic gates, we can
use one of the two existing approaches. In the first approach, one
designs reversible gates equivalent to irreversible logic gates. For
example, see Fig. 2 in which the classical irreversible gates like
NAND, AND, NOR etc are replaced by corresponding reversible gates
for reversible circuit, constructed by combination of NOT, CNOT and
CCNOT gates. In this approach after designing the equivalent gates
one can substitute each irreversible gate of a conventional circuit
by corresponding equivalent reversible gates and obtain the required
reversible circuit. This straightforward approach is used in earlier
works {[}\ref{picton}-\ref{hassan babu}] but the application of
this approach is limited since it requires an existing reversible
circuit and it can not go beyond the limits of classical computation.
Here we would like to note that all earlier efforts of designing sequential
reversible circuits {[}\ref{picton}-\ref{thapiyal2}] were limited
to this approach. For example, J. E. Rice {[}\ref{Rice}] has substituted
NOR gate (present in a conventional irreversible circuit) by CCNOT
and has used Fredkin gate to generate fanout from the clock. H. Thapiyal
{[}\ref{thapiyal2}] has substituted NAND gate by New gate and AND
gate by Fredkin gate. In the second approach one starts with a truth
table (defining the desired logic) and the gate library and then the
corresponding reversible circuits are designed by transforming the
inputs into outputs. Unidirectional and bidirectional synthesis algorithms
{[}\ref{Maslov1}] are example of this approach. In principle we have
followed this approach, but instead of following an existing synthesis
algorithm we have developed the designs by utilizing the logical symmetry
of the truth table. For example, a careful look on the truth table
of conventional SR flip flop tells us that there is an unstable condition
and as $Q$ is complement of $\overline{Q}$, so both of them can
never take same values. Let us consider two cases: (i) when clock
is high, and the last state is in Set condition and (ii) when clock
is high and the last state is in Reset condition. Now if we operate
the flip flop in Set condition (i.e. $S=1$ and $R=0$) then the state
obtained in the output in both of the above cases will be $Q^{+}=1$
and $\overline{Q}^{+}=0$. Since we obtain same result for two different
cases, it always violates bijectivity. This problem may arise in the
circuits designed by the first approach and this fact is reflected
in the state table of SR latch reported by Rice {[}\ref{Rice}]. The
truth table provided by Rice (see table V in {[}\ref{Rice}]) is not
bijective as it gives \textcolor{black}{same output (0100) for} two
different inputs (0100 and 0101). Further we would like to note that
number of feedback loops present in an irreversible sequential circuit
can not be reduced, if we adhere to the first approach.

\subsection{Gate library: Which gates should be used for the synthesis of the
reversible circuit?}

Whichever synthesis algorithm we follow, it is important to define
a gate library. The definition of the gate library (i.e. gates which
are the member of that library) is not well defined and there does
not exist any single convention. The physical complexity of gates
may not be same in two different implementation of quantum circuits.
For example, it may be easy to build an arbitrary gate 'A' in NMR
technology but it may not be that easy in superconductivity based
technology. A N-qubit quantum gate is represented by $2^{N}$x $2^{N}$
unitary matrix and product of any arbitrary number of unitary matrices
is unitary. Consequently, if we put a set of quantum gates in a black
box then an unitary matrix will represent the box and one can technically
consider it as a New gate. If we allow such construction of new gates
then any circuit block (of arbitrary size) can be reduced to a single
New gate, provided it does not contain any measurement operation.
Thus it is straightforward to observe that the use of New gate to
reduce the gate count (as it is done in {[}\ref{thapiyal2}-\ref{hassan babu}])
is an artifact. To be precise, we would like to mention that H. Thapiyal
\textit{et al} {[}\ref{thapiyal2}-\ref{thapiyal4}] has introduced
New gate, TKS gate and TSG gate to reduce the gate complexity of several
circuits. Similarly, H. M. Hasan Babu (see Fig. 7 in {[}\ref{hassan babu}])
has introduced another New gate to reduce the gate count of a full
adder circuit. Thus the gate count (gate complexity) reported in these
works {[}\ref{thapiyal2}-\ref{thapiyal4}] are misleading and consequently
we need a logical approach to construct a gate library which in turn
will help us to compare more than one circuit designs proposed for
the same purpose.

All the 1 qubit gates (set of all phase gates) along with any two
qubit gate forms an universal set of quantum gates. Since CNOT is
a two qubit gate, which have been experimentally realized by different
groups by using different techniques, it seems logical to construct
a gate library whose elements are phase gates and CNOT in case of
quantum circuit and CNOTs in case of Classical reversible circuits.
This is why we have chosen a gate library which contains NOT (N),
CNOT (C) and Toffoli (T) as its element. This particular choice of
gate library (NCT) is not only logical but also consistent with the
existing approaches {[}\ref{Maslov1},\ref{AD VOS}]. Since phase
gates and CNOT forms a set of universal gate, we can construct Toffoli
with phase gates and CNOT and that requires five gates. We have used
these facts to compare the gate complexity of our designs of sequential
circuits with the existing proposals. The comparison is done with
respect to the (NCT) gate library (as shown in Table I) and also with
respect to the universal gate library, comprising of phase gates and
CNOT (as shown in Table II).

\subsection{Why the optimized architectures are different?}

Even if we start with the same truth table and same gate library and
use the second approach of synthesis then also different logical paths
may lead to different circuits. We have followed three different logical
paths and have obtained three different architectures for every flip
flop. After designing a circuit we need to optimize it but an exact
optimization technique's time complexity $(\tau)$ is {[}\ref{Anindita}]
\begin{equation}
\tau=\mathcal{O}\left(2^{2n}n^{lm}\right)\label{eq:complexity1}\end{equation}
where, $n$ is the number of qubit line present in the circuit, $m$
is the total number gate present in the circuit and $l\leq n$, is
the number of qubit associated with the largest gate present in the
gate library. Thus in the present case when the gate library is (NCT)
then $l=3$ and the time complexity of exact optimization algorithm
is \begin{equation}
\mathcal{\tau=O}\left(2^{2n}n^{3m}\right).\label{eq:complexity2}\end{equation}
This increases exponentially with $n$ and $m$. In order to avoid
this exponential rise in time, certain approximate optimization (local
optimization) algorithms (e.g. bi-directional algorithm, template
matching algorithm) have been designed and in practice we use them.
Since the optimization algorithm is an approximate one, it may lead
to different circuit architecture but the order of gate complexities
have to be the same. This fact is clearly reflected in the Table I
and Table II given below.

\section{Proposed sequential circuits}

It is clear from the earlier approaches on reversible circuit designing
that there is no unique circuit for a particular purpose and even
if we start with the same gate library different synthesis technique
may yield different circuit architecture. More importantly, if we
obtain two different circuit architectures for the same purpose (which
satisfies the same truth table) by using two different synthesis techniques
and then apply same approximate circuit optimization algorithms (e.g.
template matching algorithm, unidirectional algorithm, bi-directional
algorithms etc.) on them it is not essential that the final circuit
will be same. \textcolor{red}{}\textcolor{black}{This fact has been
reflected in the alternative designs proposed in the present work.
In all the circuit architectures proposed in the present work} we
have used T for Toffoli (CCNOT) gate, F for Feynman (CNOT) gate and
N for NOT gate. The sum of $T_{i}$, $F_{i}$ and $N{}_{i}$ gates
present in a circuit gives us the gate count or the gate complexity
of the circuit. Here we have used a different approach and (NCT) gate
library to design reversible memory elements/ flip flops. To be precise
we have obtained three circuits for SR flip flop, JK flip flop, D
flip flop, T flip flop, Master Slave D flip flop, and Master Slave
JK flip flop. Here $D1$  refers to first feedback based design, $D2$
refers to second feedback based design and $D3$ refers to design
without feedback which, mainly uses refresh mechanism except in T
flip flop. We have included all the circuits in the present work because
of the fact that no particular design appeared as better than the
other. This fact can be seen clearly from Table I where we can see
that the third design is best to construct T flip flop but the first
design is better as far as construction of SR flip flop is concerned
and all three approaches used by us yield better results compared
to earlier works with respect to the most logical gate library (NCT).
The resultant reversible flip flop eliminates the unstable condition
found in SR flip flop and also minimizes the number of feedback loops
in the JK flip flop and Master Slave JK flip flop. The logic of each
of the circuit designed is discussed in the following subsections.

\subsection{SR Flip Flop}
\begin{figure*}[h]
\centering \scalebox{0.4}{\includegraphics{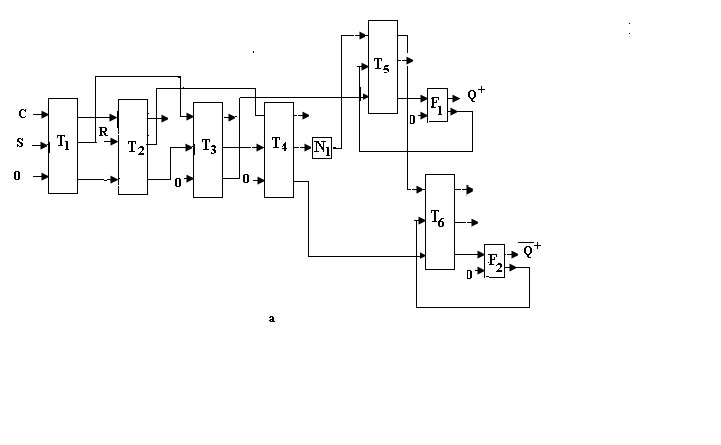}}
\begin{tabular}{cc}
\scalebox{0.3}{\includegraphics{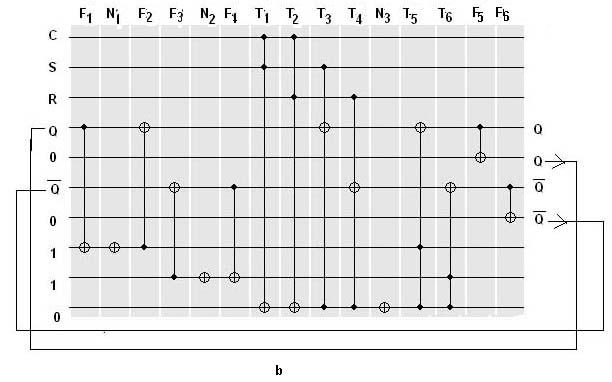}} & \scalebox{0.3}{\includegraphics{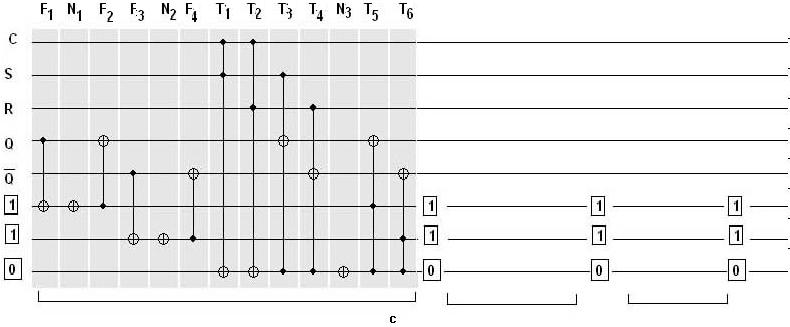}}
\end{tabular}
\caption{SR flip flop: a) design1 using feedback, b)  design2 using feedback, c)  design3 using refresh
mechanism}\label{Fig3}
\end{figure*}
\textcolor{black}{SR flip flop} or Set Reset flip flop forms the basic building block of classical flip flops.
When clock is high it gives outputs corresponding to inputs, i.e. if $R=1$, then $\overline{Q}$ = 1 and if
$S=1$, then $Q=1$. If $R=S=0$ then it holds the last state and if $R=S=1$ then it is unstable. The reversible SR
flip flop designed so far {[}\ref{picton}-\ref{thapiyal2}] have substituted AND, NAND and NOR gates present in
conventional design by their equivalent reversible gate, but we present a different logic to obtain similar
result with no unstable state. Fig. 3 shows our designs of SR flip flop. When clock is high and either inputs
are high then flip flop is Set or Reset but when both inputs are high or low then it retains its last state also
when clock is low then also it retains its last state.

In Fig. 3a, when clock is high, then $T{}_{1}$ and $T_{2}$ will compare the inputs. This implies that
$(C.S)\oplus(C.R)=1$ only when clock ($C$) and one of the inputs (i.e. either $S$ or $R$) are high. The required
outputs $(Q^{+}$ and $\bar{Q}^{+})$ corresponding to this particular situation (i.e. when $(C.S)\oplus(C.R)=1$)
are obtained from target bits of $T{}_{3}$ and $T_{4}$ respectively. But the situation is different if the
result is low $($ i.e. $(C.S)\oplus C.R)=0$$)$. This happens in three cases: (i) When clock is low, (ii) when
clock is high and both the inputs are low and (iii) when clock is high and both the inputs are high. If we use a
conventional SR flip flop circuit made of NAND\textbackslash{}NOR gate then we obtain unstable state in output
in case (ii)\textbackslash{}case (iii) above. Here we want to go beyond the domain of classical irreversible
circuit and design a reversible SR flip flop circuit free from unstable condition. To do so we have added a NOT
gate (on the second qubit line of the output of $T_{4}$) which yields $\overline{(C.S)\oplus(C.R)}$.
Consequently, the target bits of $T{}_{5}$ and $T{}_{6}$ will give the output which will be the last state or
the previous state. Here we have described the logical path of the first design (i.e $D1$) in detail because
this design has less number of gates compared to other two designs ($D2$ and $D3$) obtained by us by using other
logical paths. In Fig. 3b, $D2$ design is shown which first copies the last state of $Q$ and $\overline{Q}$ in
$8^{th}$ and $9^{th}$ qubit lines from $4^{th}$ and $6^{th}$ qubit lines and then sets it to zero so that it can
store new output values ($Q^{+}$ and $\overline{Q}^{+}$). This implies that whenever we need to compare last
values ($Q$ and $\overline{Q}$ ) we can obtain it from $8^{th}$ and $9^{th}$ qubit lines. It further follows
similar logic as in $D1$ but has higher gate count. In Fig. 3c, $D3$ design is similar to $D2$ but only
difference is that it does not uses any feedback, rather it refreshes the last three qubit lines before any
operation.
\begin{figure} [h]
\centering \scalebox{0.4}{\includegraphics{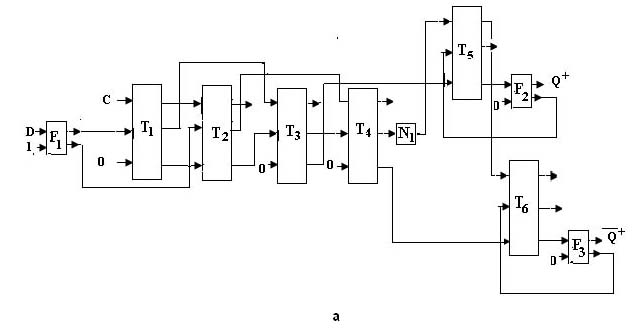}}
\begin{tabular}{cc}
\scalebox{0.4}{\includegraphics{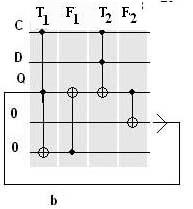}} & \scalebox{0.4}{\includegraphics{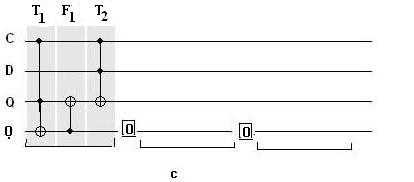}}
\end{tabular}
\caption{D flip flop: a) design1 using feedback, b) design2 using feedback, c) design3 using refresh mechanism
}\label{Fig4}
\end{figure}

\subsection{D Flip Flop}

In conventional irreversible logic, D flip flop \textbf{}is built
by using a NOT gate between the two inputs of SR flip flop. Consequently,
output Q follows D when clock is high and it stores the data otherwise.
Fig. 4a shows reversible realizations of D flip flop by using similar
idea. \textcolor{black}{In Fig. 4b when clock is high, value of $Q$
is copied in $5^{th}$ qubit line and thus the next operation (i.e.
$Q\oplus Q$) reduces the $3^{rd}$ qubit to zero and after it copies
D by using the last Toffoli gate ($T_{2}$). But if the clock is low
then the $3^{rd}$ qubit ($Q$) will retain its last value.} In Fig.
4c feedback is avoided, thus there is no need of ancilla bit ($5^{th}$
qubit line) and the $4^{th}$ qubit line is refreshed after every
operation.

\subsection{JK flip flop}
\begin{figure}[h]
\centering
\begin{tabular}{cc}
\scalebox{0.4}{\includegraphics{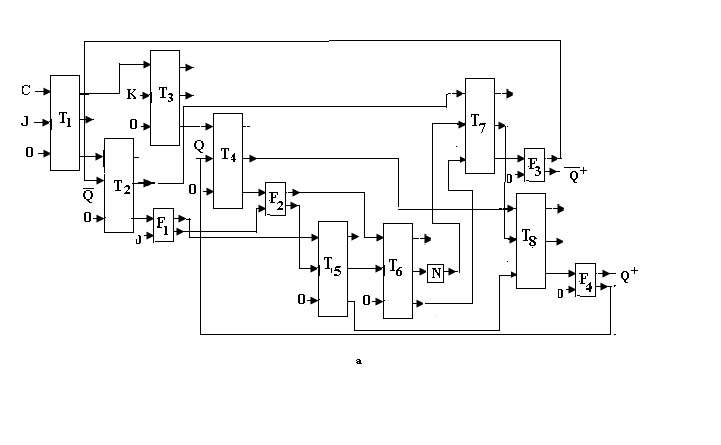}} & \scalebox{0.3}{\includegraphics{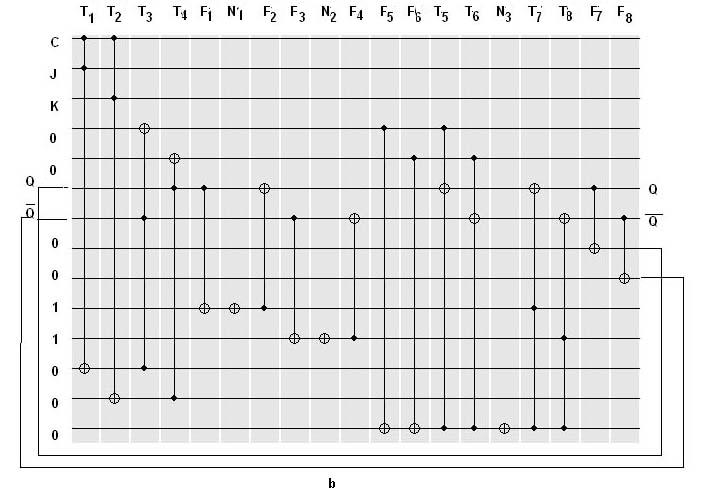}}
\end{tabular}
\caption{JK flip flop: a) design1 using feedback and b) design2 using feedback }\label{Fig5}
\end{figure}
Classically JK flip flop is built from SR flip flop. The outputs of SR flip flop and inputs of JK flip flop
drives SR block such that when the last state $\overline{Q}$ and $J=1$ then $Q^{+}=1$$ $and when last state $Q$
and $K=1$ then $\overline{Q}^{+}=1$. In case of similar inputs, $J=K=1$ output toggles and for $J=K=0$ output
holds the last state. Fig. 5 shows JK flip flop built on this logic. In Fig. 5a, $T_{1}$ and $T_{3}$ will give
the values of J and K only when clock is high. The next step is to obtain the values of S and R depending
\textcolor{black}{upon val}ues of $J$ and $K$ and feedback values of $\overline{Q}$ and $Q$. The result is
obtained from the target bits of $T_{2}$ and $T_{4}$ respectively. The rest is SR block, its $T_{1}$ and $T_{2}$
is reduced here to CNOT gates ($F_{1}$ and $F_{2}$). In this design feedback values Q and $\overline{Q}$ are
compared four times i.e. four gates $($$T_{2},\, T_{4},\, T_{7}$ and $T_{8})$ uses feedback values as controls
but external feedback is used twice. In $D3$ we have removed the copy gates from Fig. 5b and last five bit lines
have been refreshed to initially set values.

\subsection{T flip flop}
\begin{figure}[h]
 \centering \scalebox{0.4}{\includegraphics{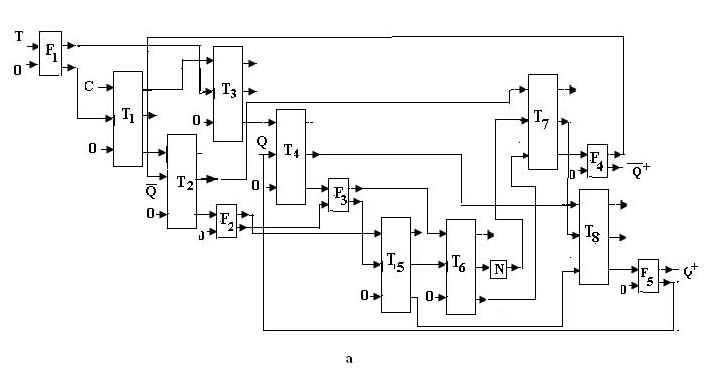}}
\begin{tabular}{cc}
\scalebox{0.4}{\includegraphics{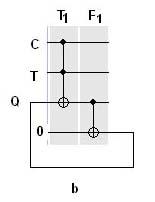}} & \scalebox{0.4}{\includegraphics{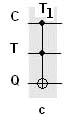}}
\end{tabular}
\caption{T flip flop: a) design1 using feedback, b) design2 using feedback and c) design3} \label{Fig6}
\end{figure}
As the name suggests, this flip flop circuit used to toggle the output when input is high $(1)$ and retains the
output when input is low $(0)$, thus it does two operation, it either holds the last state or toggles the
output. Essentially, it has a logical symmetry with Controlled NOT kind of operation. In fact, it can be made by
joining J and K inputs from JK flip flop. In the first design, which appeared as the best approach for designing
SR flip flop, we can achieve the goal of designing JK flip flop simply by adding a CNOT gate in JK flip flop
(see Fig. 6a). Similar operation can also be achieved by an alternative logical approach (as shown in Fig. 6b)
by using CCNOT and CNOT gates. Here a CNOT copies the output and send it to CCNOT, which operates when clock is
high (toggles) and when clock is low then it retains the last state. As far as gate count is concerned this
second design is much better than the first one. But still it has not been successfully exploited the logical
symmetry of the T flip flop with Controlled NOT kind of operation. Now, since a qubit line can always be used to
hold the last state, thus we can represent T flip flop by a Toffoli gate (as in Fig. 6c). Therefore, to design a
reversible T flip flop neither feedback nor the refreshment mechanism is required. This has a sharp contrast
with the classical design of T flip flop.

\subsection{Master Slave D flip flop and Master Slave JK flip flop}
\begin{figure}[h]
\centering \scalebox{0.4}{\includegraphics{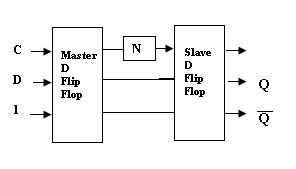}}
 \caption{MasterSlave D FlipFlop }\label{fig6}
\end{figure}
\begin{figure}[h]
\centering \scalebox{0.4}{\includegraphics{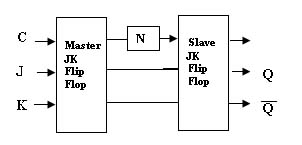}} \caption{MasterSlave JK FlipFlop }\label{fig7}
\end{figure}
Master Slave flip flops are formed by joining two similar blocks. The former block is driven by Clock $(C)$ and
later block by inverse of clock $(\overline{C})$. Apparently the second block behaves as Slave and follows the
Master, i.e the first block. Master Slave D flip flop (Fig. 7) is made by joining two D flip flops where the
later is driven by $\overline{C}$. \textcolor{black}{Master block will have 10 gates while Slave block will have
9 gates since it does not require $F_{1}$ gate (see Fig. 4a) and there will be a NOT gate to obtain
$\overline{C}$ from $C$. Thus its gate count in design 1 of Master Slave flip flop will be 20. In design 2 and
3, Master and Slave block will be similar in construction i.e. have equal number of gates and will also have a
NOT gate between them. Similarly Master Slave JK flip flop (Fig. 8) is made by adding two JK blocks and driving
the later by $\overline{C}$. The M}aster block behaves as JK flip flop of $D1$ and has 13 gates, its output are
given to Slave block that is similar to that of SR block of $D1$ along with two CNOT gates for fan out and in
total has 25 gates.

\section{Comparison }

Since the earlier designs of reversible circuits use different gate
libraries. For the purpose of comparison of circuit complexity of
our proposals with the existing proposals we have \textcolor{black}{followed
the steps given below} steps:

\begin{enumerate}
\item An equivalent circuit (using NCT gate library) is constructed for
each non-NCT gates used by Picton {[}\ref{picton}], Rice {[}\ref{Rice}]
and Thapiyal {[}\ref{thapiyal2}]. This has been done with the help
of uni-directional algorithm and bi-directional algorithm. Normally
it is found that the bi-directional algorithm provides better result
(i.e. lesser gate complexity).
\item The equivalent circuits constructed by the above techniques are then
optimized with the help of template matching algorithm, moving rule
and deletion rule {[}\ref{Maslov1}]. For example, the New gate introduced
in {[}\ref{hassan babu}] requires 7 NCT-gates, the New gate introduced
in {[}\ref{Thapiyal}] requires 5 NCT-gates and Fredkin gate requires
3 NCT-gates.
\item Once the optimized circuits equivalent to non-NCT gates are obtained,
they are replaced in the original circuits of Thapiyal, Rice and Picton.
Thus the essential logic remained same.
\item After obtaining the NCT equivalent and logic conserving circuits of
earlier proposals, the optimization techniques (i.e. template matching
algorithm, moving rule and deletion rule) are applied once again on
the whole circuit to obtain optimized, NCT equivalent and logic conserving
circuit of the earlier proposals. Number of NCT gates present in \textcolor{black}{these
circuits is counted} and this count is considered as the complexity
of the circuit.
\end{enumerate}
Finally, we have compared the gate complexity of different designs
in {[}\ref{Rice}, \ref{thapiyal2}] with the corresponding proposals
of the present work (see Table I below) and have found that the present
proposals have lower gate complexity as well as the lower number of
garbage output compared to the earlier proposals, which satisfies
all the requirement of bijectivity.

\begin{center}
~
\par\end{center}

\begin{center}
\begin{tabular}{|c|c|c|}
\hline
&
No. of Gates &
No. of Garbage outputs\tabularnewline
\multicolumn{1}{|c|}{\begin{tabular}{c}
\tabularnewline
\hline
\multicolumn{1}{c}{SR F/F}\tabularnewline
\hline
D F/F\tabularnewline
\hline
JK F/F\tabularnewline
\hline
T F/F\tabularnewline
\hline
MS D F/F\tabularnewline
\hline
MSJK F/F\tabularnewline
\end{tabular}}&
\begin{tabular}{cccccc}
&
Rice&
Thapiyal&
$D1$&
$D2$&
$D3$\tabularnewline
\hline
&
9&
18&
9&
15&
13\tabularnewline
\hline
&
-&
23&
10&
4&
3\tabularnewline
\hline
&
-&
26&
13&
19&
17\tabularnewline
\hline
&
-&
26&
14&
2&
1\tabularnewline
\hline
&
17&
-&
20&
9&
7\tabularnewline
\hline
&
-&
54&
23&
37&
33\tabularnewline
\end{tabular}&
\begin{tabular}{cccccc}
&
Rice&
Thapiyal&
$D1$&
$D2$&
$D3$\tabularnewline
\hline
&
7&
8&
6&
6&
6\tabularnewline
\hline
&
-&
8&
6&
3&
3\tabularnewline
\hline
&
-&
12&
10&
10&
10\tabularnewline
\hline
&
-&
12&
10&
2&
2\tabularnewline
\hline
&
12&
-&
11&
5&
5\tabularnewline
\hline
&
-&
21&
15&
18&
18\tabularnewline
\end{tabular}\tabularnewline
\hline
\end{tabular}
\par\end{center}

\begin{center}
~
\par\end{center}

Table I: A comparison table of gate complexity and garbage outputs
reported in {[}\ref{Rice},\ref{thapiyal2}] with the present proposals.
The comparison is done with respect to NCT gate library.

~

It is well known that the set of any two qubit gate and all possible
one qubit gate forms a set of universal gate. The library used till
now is not universal and it may be tempting to see what happens if
one uses such a gate library. Since the two choice of two qubit gate
is not unique so we can have many such universal set, but following
the earlier logic we can conclude that the set of CNOT and all one
qubit operations (or CNOT and a Phase gate in general%
\footnote{Remember that now the number of element in the set is infinite and consequently it is impossible to
design a polynomial time algorithm which can deterministically optimize a quantum circuit. What ever be our
choice of gate library, unless the choice of gate is restricted to a subset of the universal set we can not
circumvent the problem associated. In that sense finite libraries like (NCT) or (NCT and
Hadamard) are good choices.%
}) forms the best choice for the alternative gate library. In such
case gate complexity of a Toffoli gate become 5 and if we consider
this fact while circuit complexities of the circuits designed so far
then Table I reduces to Table II below.

~

\begin{center}
\begin{tabular}{|c|c|}
\hline
&
No. of Gates \tabularnewline
\multicolumn{1}{|c|}{\begin{tabular}{c}
\tabularnewline
\hline
SR F/F\tabularnewline
\hline
D F/F\tabularnewline
\hline
JK F/F\tabularnewline
\hline
T F/F\tabularnewline
\hline
MS D F/F\tabularnewline
\hline
MSJK F/F\tabularnewline
\end{tabular}}&
\begin{tabular}{cccccc}
&
Rice&
Thapiyal&
$D1$&
$D2$&
$D3$\tabularnewline
\hline
&
29&
32&
33&
39&
37\tabularnewline
\hline
&
-&
55&
34&
12&
11\tabularnewline
\hline
&
-&
58&
45&
51&
49\tabularnewline
\hline
&
-&
58&
46&
6&
5\tabularnewline
\hline
&
53&
-&
68&
19&
17\tabularnewline
\hline
&
-&
126&
79&
93&
89\tabularnewline
\end{tabular}\tabularnewline
\hline
\end{tabular}
\par\end{center}

\begin{center}

\par\end{center}

\begin{center}
\textcolor{black}{~}
\par\end{center}

\textcolor{black}{Table II: A comparison table of gate complexity
reported in {[}\ref{Rice},\ref{thapiyal2}] with the present proposals.
The comparison is done with respect to an universal gate library which
contains all one qubit gates (all phase gates) and CNOT as its element.
Here we count complexity of CCNOT gate as 5.}

~

It is interesting to note that the advantages of our design over the
earlier proposals and all other conclusions remained same. This approach
of counting circuit complexity by considering complexity of Toffoli
as 5 is consistent with the earlier works {[}\ref{Maslov1}].

\section{Conclusions}

In section 2 of the present work, we have addressed the conceptual
issues related to the designing and optimization of reversible circuits
in general with a special attention towards the issues related to
the designing of reversible sequential circuits. In this section we
have shown that it is required to define an acceptable gate library
and a good choice for that can \textcolor{black}{be NCT} gate library.
Further, it has been shown that the advantage in gate count obtained
in some of the earlier proposals by introduction of \textcolor{black}{New
gates or unconventional gates (such as TSG and TKS gates)} is an \textcolor{black}{artifact}
and if it is allowed then every circuit block (unless there is a measurement)
can be reduced to a single gate. The important conclusions of section
2 have been used in the next section to design \textcolor{black}{locally
optimized reversible} circuits for SR flip flop, JK flip flop, D flip
flop, T flip flop, Master Slave D flip flop and Master Slave JK flip
flop. In section 4 we have seen that our \textcolor{black}{designs
not only} \textcolor{red}{}\textcolor{black}{overcome} the unstable
condition i.e. it always holds the last state \textcolor{black}{when
inputs are similar} but also uses lesser number feedback loops (i.e.
only 2 feedback loops against 4 in conventional JK flip flop). In
the third design feedback loops are replaced by application of refresh
mechanism on some qubit lines which can also be reduced (in SR flip
flop it can be minimized to 1 refresh operation on qubit line) but
it leads to increase of gate count and consequently it has been avoided
in the present work. From the comparison tables (Table I and Table
II), it is clear that the none of the earlier proposals have lesser
gate complexity and lesser number of garbage bits compared to the
present proposals.

The sequential circuits described here uses only NCT gates and there exist several proposals for realization of
CNOT and CCNOT gates using CMOS based technology {[}\ref{vasudevan}-\ref{AD VOS}]. Thus it is technically
possible to construct the proposed reversible sequential circuits with the help of conventional CMOS based
technology. Consequently, we can easily build \textcolor{black}classical reversible memory element. But the
implementation of the present work is not limited to classical domain, this is because of the fact that we can
also implement the proposed circuits in quantum domain with the help of NMR or quantum dot or optical
implementation {[}\ref{the:cerf-teleportation}]. If one aims to provide optimized reversible circuits for all
the useful component of a classical computer and then this work along with the proposal of {[}\ref{AD VOS}] will
help him to provide a complete architecture for a classical reversible computer. Since it will be free from the
problem of decoherence and scalability it seems more practical and easy to built than a real scalable quantum
computer.

~

\textbf{Acknowledgement:} A. P. thanks, Department of Science and
Technology, India, for the partial financial support provided through
the project no. SR\textbackslash{}FTP\textbackslash{}PS-13\textbackslash{}2004.

\end{document}